\documentclass[a4paper]{jpconf}
\usepackage{graphicx}
\usepackage{breqn}
\usepackage{hyperref}
\hypersetup{colorlinks = false}
\begin{document}
\title{Report on Thermal Neutron Diffusion Length Measurement in Reactor Grade Graphite Using MCNP and COMSOL Multiphysics}

\author{S. R. Mirfayzi}

\address{School of Physics and Astronomy
University of Birmingham
Birmingham
B15 2TT
United Kingdom}

\ead{srm105@bham.ac.uk}

\pagestyle{plain}
\pagenumbering{arabic}

\graphicspath{{/}{/}}
\bibliographystyle{iopart-num}

\begin{abstract}
Neutron diffusion length in reactor grade graphite is measured both experimentally and theoretically. The experimental work includes Monte Carlo (MC) coding using 'MCNP' and Finite Element Analysis (FEA) coding suing 'COMSOL Multiphysics' and Matlab. The MCNP code is adopted to simulate the thermal neutron diffusion length in a reactor moderator of 2m x 2m with slightly enriched uranium ($^{235}U$), accompanied with a model designed for thermal hydraulic analysis using point kinetic equations, based on partial and ordinary differential equation.    
The theoretical work includes numerical approximation methods including transcendental technique to illustrate the iteration process with the FEA method. Finally collision density of thermal neutron in graphite is measured, also specific heat relation dependability of collision density is also calculated theoretically, the thermal neutron diffusion length in graphite is evaluated at $50.85 \pm 0.3cm$ using COMSOL Multiphysics and $50.95 \pm 0.5cm$ using MCNP. Finally the total neutron cross-section is derived using FEA in an inverse iteration form.   
\end{abstract}

\section{Introduction}
This work demonstrates an analytic approach accompanied with models of Finite Element Analysis (FEA) and Monte Carlo (MC) with an experimental measure on neutron cross-section and slowing down process. In MC approach Monte Carlo N-Particle Transport Code (MCNP) is used to simulate the simplified version of reactor moderation process. Similarly in FEA the moderator modelled (Assuming a symmetrical distribution) using point kinetic equations, based on partial and ordinary differential equation in software package.

\section{Theoretical Calculations}
\begin{figure}[!htbp]
  \begin{center}
    \leavevmode
      \includegraphics[height=3.5in]{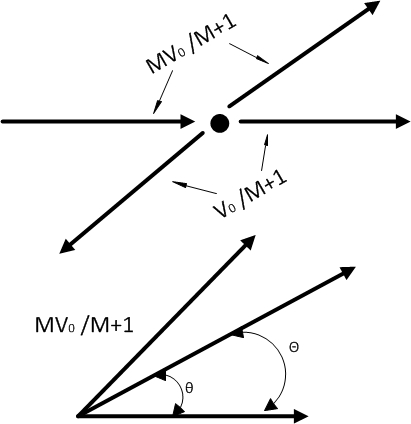}
    \caption[The Collision in Centre-of-Mass System.] {The Collision in Centre-of-Mass System:If a neutron with initial velocity $V_0$ collides with a nucleus of mass M at rest, then in the  Centre of Mass (CoM)system the initial velocity is $M V_0 /M+1$ after collision. The momentum of of neutron and the nucleus will be equal to magnitude oppositely directed vector.Here $\theta$ is the deflection angle and $\Theta$ is angle on the final velocity $v$.}
    \label{VelocityVecI}
  \end{center}
\end{figure}

\begin{figure}[!htbp]
  \begin{center}
      \includegraphics[height=2.5in]{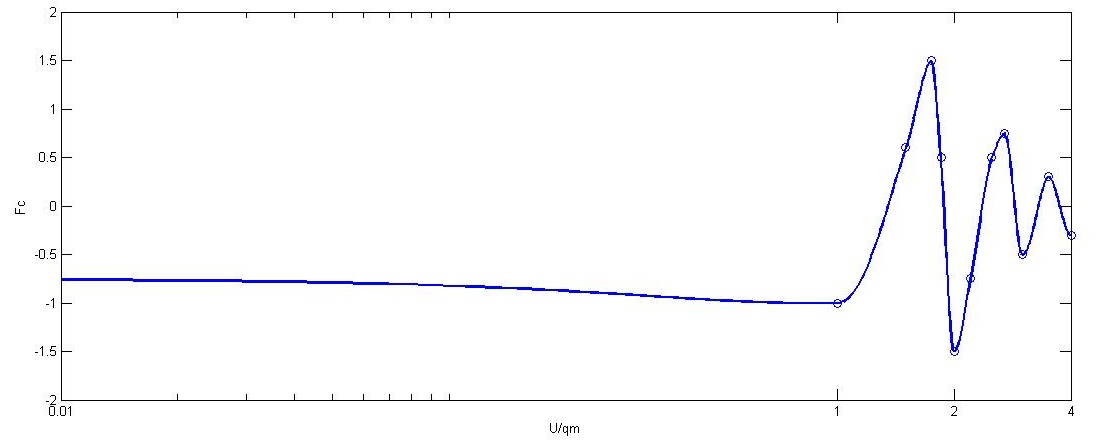}
    \caption[The Maximum Logarithmic Energy Loss vs. Collision Density] {The Maximum Logarithmic Energy Loss vs. Collision Density}
    \label{maxLogarithmic}
  \end{center}
\end{figure}

\begin{figure}[!htbp]
  \begin{center}
    \leavevmode
      \includegraphics[height=2.67in]{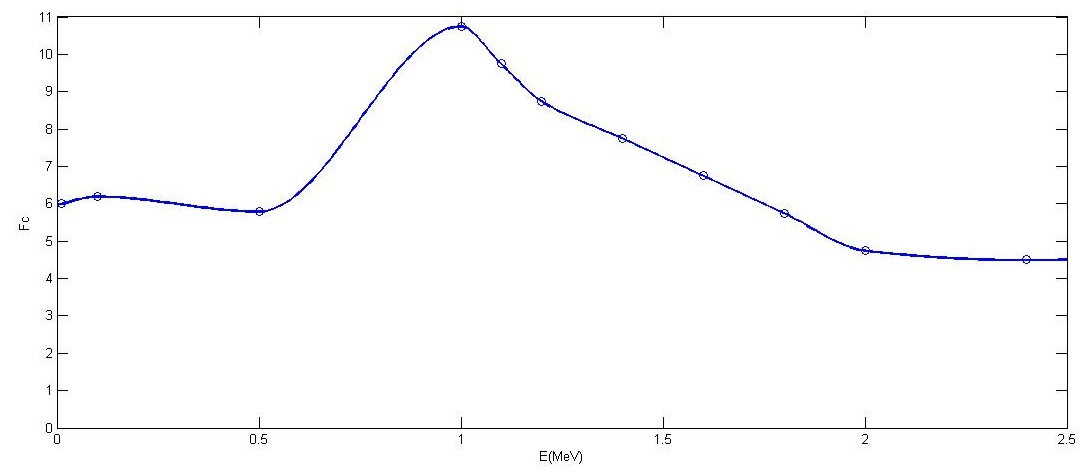}
    \caption[The Collision Density vs. Neutron Energy in Graphite] {The Collision Density vs. Neutron Energy in Graphite}
    \label{GraphiteUqm}
  \end{center}
\end{figure}

Having the number of particles found in a volume element dr where $dr=dxdydz$ at $r$ with a vector with solid angle $d \Omega $ at $\Omega$ be donated by \cite{MarshakI}:
\newline
\begin{equation}
\label{eq1}
N(r,\Omega,t)dr d \Omega
\end{equation}
\newline
Therefore can have:
\newline
\begin{equation}
\label{eq2}
\frac{dN}{dt} = -NV \sigma + \int N(r, \Omega',t)V \sigma_s f(\Omega.\Omega')d \Omega'+ S(r, \Omega ,t)
\end{equation}
\newline
Where the first term $\frac{dN}{dt}$ donates the number of particles present in given volume (particle density) and second term $-NV \sigma$ represents the total number of particles removed from the given volume by scattering and capture. $\sigma$ is representing the total cross-section. The third term represents the total number of particles scattered into the given volume, and $f(\Omega.\Omega')$ represents the relative probability of scattering through an angle whose cosine is $\Omega.\Omega'$, where $\Omega'$ is a unit vector in the direction of the initial velocity and $\Omega$ is unit vector in the final direction. Finally $S(r, \Omega ,t)$ is the external source term available in the system and is given by:
\newline
\begin{equation}
\label{eq3}
N(r,\Omega,t)dr d \Omega = N(Z, \varphi)dZ d \varphi d \phi
\end{equation}
\newline 
Where $\varphi$ is the cosine of the velocity vector in the Z direction and $\phi$ is the longitude of velocity vector. 
\newline
\begin{figure}[!htbp]
  \begin{center}
    \leavevmode
      \includegraphics[height=2.8in]{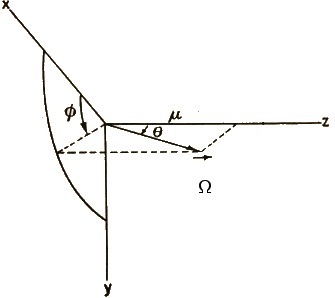}
    \caption[The Velocity Vector] {The Velocity Vector: Where $\varphi$ is the cosine of the velocity vector in the Z direction and $\phi$ is the longitude of velocity vector and $\Omega$ is unit vector in the final direction. }
    \label{VelocityVec}
  \end{center}
\end{figure}
\newline 
Now an assumption can be made such:
\newline
\begin{equation}
\label{eq4}
N_0(Z)=2 \pi \int_{-1}^{+1} N(Z, \varphi) d \varphi
\end{equation}
\newline 
Rewrite the equation \ref{eq1} as:
\newline
\begin{equation}
\label{eq5}
V_{\varphi} \frac{\delta N (Z, \varphi)}{dZ} = -N(Z, \varphi) V \sigma + \int N(Z, \varphi') V \sigma_s f(\varphi_0) d\Omega + S(z)
\end{equation}
\newline 
Where $\varphi_0$ is the cosine of the angle between initial and final velocities and it can be found by:
\newline
\begin{equation}
\label{eq6}
cos \theta cos \theta' + sin \theta sin \theta' cos(\phi -\phi') 
\end{equation}
\newline 
Or using:
\newline
\begin{equation}
\label{eq7}
\varphi_0= \varphi \varphi'+  \sqrt{1- \varphi^2} \sqrt{1- \varphi^{'2}}  cos(\phi -\phi') 
\end{equation}
\newline 
Now if the collision function of $F(\varphi_0)$ expanded in spherical harmonics:
\newline
\begin{equation}
\label{eq8}
F(\varphi_0)=\sum_{0}^{\infty } \frac{2l+1}{4\pi}F_1P_1 (\varphi_0)
\end{equation}
\newline 
With $F_1= \int f(\varphi_0) P_1(\varphi_0) d \Omega $. Using similar expansion the phase density function will be:
\newline
\begin{equation}
\label{eq9}
N(Z, \varphi)=\sum_{0}^{\infty } \frac{2l+1}{4\pi} N_1(Z) P_1 (\varphi)
\end{equation}
\newline 
Where $N(Z, \varphi)= \int N(Z, \varphi) P_1 (\varphi) d \Omega $, with assumption that $N(Z,\varphi)$ is isotropic, three conditions must be satisfied all the times: first, it is far from the source (equal to Mean Free Path (MFP)), second, it is far from the boundaries; third, the probability of capture is small compared to probability of scattering. Having all the conditions satisfied, the following can be assumed:
\newline
\begin{equation}
\label{eq10}
N(Z, \varphi) \cong \frac{1}{4 \pi} ( N_0(Z) + 3 \varphi N_1(Z) ) 
\end{equation}
\newline 
Where the second term in the bracket donates the particle flux (J). Here $N_1= \int \varphi N(Z, \varphi)d \Omega =J/V$. For simplicity we choose our unit such that $V=1$ and $\sigma =1 $, hence:
\newline
\begin{equation}
\label{eq11}
1-f= \frac{\sigma_s}{\sigma}
\end{equation}
\newline 
Now the Boltzmann equation takes the form of:
\newline
\begin{equation}
\label{eq12}
N \frac{dN}{dZ}= - N+(1-f) \int N(Z, \varphi')f(N_0) d \Omega' + S(Z)
\end{equation}
\newline 
By integrating the equation over all possible angles $(d \Omega)$ we have:
\newline
\begin{equation}
\label{eq13}
 \frac{dN_1}{dZ}= - N_0 +(1-f)N_0 + 4 \pi S(Z)
\end{equation}
\newline
$f$ is normalized in such a way that $\int f(\Omega. \Omega')d \Omega'= \int f(\Omega. \Omega)d \Omega =1 $, hence going back to Eq. \ref{eq9} for the case $l=0$ we have:
\newline
\begin{equation}
\label{eq14}
F_0= \int f(\varphi_0) d \Omega=1
\end{equation}
\newline
Hence by integrating over all angles and Multiplying by $\varphi$ we have:
\newline
\begin{equation}
\label{eq15}
 \frac{1}{3} \frac{dN_0}{dZ}= -N_1 +(1-f)F_1N_1
\end{equation}
\newline
Now the second order differential equation gives:
\newline
\begin{equation}
\label{eq16}
- \frac{1}{3 (1-f_1} \frac{d^2 \varphi_0}{dZ^2}=-f N_0 +S_0(Z)
\end{equation}
\newline
This also can be written as:
\newline
\begin{equation}
\label{eq17}
\bigtriangledown^2 N_0- \frac{1}{L^2} N_0 + \frac{1}{D} S_0=0
\end{equation}
\newline
Eqs. \ref{eq16} and \ref{eq17} are known as diffusion equation. Here $L$ is diffusion Length abd D is diffusion coefficient and it is equal to $\frac{1}{3} \frac{\lambda_s}{1-f_1}= \frac{\lambda_{tr}}{3} V$, where $\lambda_s$ and $\lambda_{tr}$ are the scattering and transport mean free path. $\lambda_{tr}$ can be calculated from:
\newline
\begin{equation}
\label{eq18}
\lambda_{tr}= \frac{\lambda_s}{1- (cos \theta)_{av}}
\end{equation}
\newline
and $(cos \theta)_{av}$ is equal to $ \int f (\varphi_0) \varphi_0 d \Omega=f_1$. Also $L^2$ can be measured using following relation:
\newline
\begin{equation}
\label{eq19}
L^2=\frac{\lambda_c \lambda_{tr}}{3}
\end{equation}
\newline
Here $\lambda_c$ is the capture mean free path. 
\newline
\newline
{\bf MAXIMUM ENERGY LOSS}
\newline
If a neutron with initial velocity $V_0$ collides with a nucleus of mass M (at rest), then in the  Centre of Mass (CoM) system, the initial velocity is $M V_0 /M+1$ after collision. The momentum of of neutron and the nucleus will be equal to magnitude oppositely directed vector. Figure \ref{VelocityVecI} demonstrates the collision in CoM system.    
\newline 
As demonstrated in Fig. \ref{VelocityVecI} the $\theta$ is the deflection angle and $\Theta$ is angle on the final velocity $v$. The $v^2$ in this case is given by:
\newline
\begin{equation}
\label{eq20}
\frac{Mv_0}{M+1}cos \theta+ \frac{v_0}{M+1}=vcos\Theta
\end{equation}
\newline
\newline
\begin{equation}
\label{eq21}
(\frac{Mv_0}{M+1} )^2 + (\frac{v_0}{M+1} )^2 - \frac{2Mv^{2}_{0}}{M+1} cos \theta=v^2 
\end{equation}
\newline
so
\newline
\begin{equation}
\label{eq22}
cos \theta = 1- \frac{(M+1)^2}{2M} 1 - \frac{v}{v_0})^2
\end{equation}
\newline
since $u=log \frac{E_0}{E}$
then:
\newline
\begin{equation}
\label{eq23}
cos \theta = 1- \frac{(M+1)^2}{2M} 1 - e^{-u}
\end{equation}
\newline
now differential cross-section gives:
\newline
\begin{equation}
\label{eq24}
\frac{dcos \theta}{du}= - \frac{(M+1)^2}{2M}  e^{-u}
\end{equation}
\newline
Hence:
\newline
\begin{equation}
\label{eq25}
cos \Theta = - \frac{(M+1)^2}{2} e^{\frac{-u}{2}} - \frac{M-1}{2}e^{\frac{u}{2}}
\end{equation}
\newline
Therefore the maximum logarithmic energy loss can be calculated from:
\newline
\begin{equation}
\label{eq26}
q_M=log (\frac{M+1}{M-1})^2
\end{equation}
\newline
The $q_M$ is at most when $\Theta=\pi$. Now going back to the problem we can redefine the collision density function as:
\newline
\begin{equation}
\label{eq27}
F( \varphi_0,u)= \frac{(M+1)^2}{8 \pi M}e^{-u} \times \delta ( \varphi_0 - ( \frac{(M+1)}{2} e^{\frac{-u}{2}} - \frac{M-1}{2} e^{\frac{u}{2}} )
\end{equation}
\newline
The term $\frac{(M+1)^2}{8 \pi M}e^{-u}$ is the normalization constant chosen to satisfy $\int d \Omega \int du f(\varphi_0, u) =1$ and $\delta$ is the Dirac $\delta$ function. So that $\delta(x-a)=0$ when $x \neq 0$ and $\int d (x-a) F(x)dx=F(a)$. Now the average logarithmic loss $(\xi)$ can be calculated from:
\newline
\begin{equation}
\label{eq28}
\xi = 1- \frac{(M+1)^2}{4M} q_m e^{-q_m} 
\end{equation}
\newline
and $(cos)_av$ is $2/3M$.
\newline
\newline
{\bf Energy Distribution of Slowed Down Neutrons}
\newline
{\bf I. Stationary Case}
\newline
The average collision density per unit time, with logarithmic energy intervals is given by:
\newline
\begin{equation}
\label{eq29}
\psi_0(u)= \int_{0}^{u} du' \psi_0(u') h(u') f_0 (u-u') +\delta (u)
\end{equation}
\newline
where $f_0(u)$ is $(M+1)^2 e^{-u}/4m$ for $u \leq q_m $ and it is zero otherwise. In stationary case the total number of neutron produced is unity per unit in this case, i.e. for $M=12$, $f+0(u)$ becomes $3.5e^{-u}$. Hence the equation \ref{eq29} becomes:
\newline
\begin{equation}
\label{eq30}
\psi_0(u)= \int_{u-q_m}^{u} du' \psi_0(u') h(u') 3.5e^{-(u-u')} +\delta (u)
\end{equation}
\newline
where $q_m=0.72$ 
\newline
\newline 
{\bf II. Time-dependent Case}
\newline
The time dependent when there is no absorption in the system and source strength is unity and is given by: 
\newline
\begin{equation}
\label{eq30}
\frac{l(u)}{v} \frac{d \psi_0}{\delta t}+ \psi_0(u,t)= \int_{0}^{u} du' \psi_0(u',t)e^{-(u-u')} +\delta(u) \delta(t)
\end{equation}
\newline
where $l(u)$ is the mean free path and if the mean free path is constant, the Laplacian form of the equation for $M \neq 1$ becomes:
\newline
\begin{equation}
\label{eq31}
1+ \frac{sl_0}{v} \phi_0(u,s)= \int_{0}^{u} du' \phi_0(u',s) f_0 (u-u') +\delta(u)
\end{equation}
\newline
now:
\newline
\begin{equation}
\label{eq32}
\phi(w,s)=  \frac{2}{(1-r^2)w^2}\int_{rw}^{w} dw' \frac{w' \phi(w',s)}{(1+w')}
\end{equation}
\newline
Eq. \ref{eq32} applies for $u>q_m$ where $w=l_0s/v$, $r=M-1/M+1$ and $\phi(w,s)=(1+w)\phi_0(u,s)$, so that the mean free path is proportional to velocity.
\newline
\newline
{\bf III. Rigorous Numerical Solution}
\newline 
The slowing down process is not an easy approach, therefore a more discrete form of solution also could be defined using:
\newline
\begin{equation}
\label{eq33}
F(E)= \int_{E}^{\infty } \sum _S(E' \rightarrow E) \phi (E')dE'+ \delta(u)
\end{equation}
\newline
Where $\sum _S(E' \rightarrow E)$ is the scattering term between energies $E'$ and $E$. Recalling $\sum_S$:
\newline
\begin{equation}
\label{eq34}
\sum _S(E' \rightarrow E) = \sum _S(E') P(E' \rightarrow E)
\end{equation}
\newline
Where $P(E' \rightarrow E)$ is the probability of collision happens between $E'$ and $E$. It can be defined by:
\newline
\begin{equation}
\label{eq35}
P(E' \rightarrow E).E'(1- \alpha)=1
\end{equation}
\newline
Hence:
\newline
\begin{equation}
\label{eq36}
P(E' \rightarrow E)= \frac{1}{(1- \alpha)E'}
\end{equation}
\newline
Now:
\newline
\begin{equation}
\label{eq37}
\sum _S(E' \rightarrow E)= \frac{\sum _S(E')}{(1- \alpha)E'}
\end{equation}
\newline
For $M=1$ the collision is defined as:
\newline
\begin{equation}
\label{eq38}
F(E)= \int_{E}^{\infty } \frac{\sum _S(E')}{E'} \phi (E) dE' +\delta(E)
\end{equation}
\newline
Also the solution with capture process:
\newline
\begin{equation}
\label{eq39}
F_c(E)= \frac{\sum _S(E_0)}{\sum _t(E)} \frac{S_0}{E} exp(- \int_{E}^{E_0} \frac{\sum _S(E')}{\sum _t(E')} \frac{dE'}{E'}  ) 
\end{equation}
\newline
Also For $M \neq 1$ the collision density can be found:
\newline
\begin{equation}
\label{eq40}
F(E)= \int_{E/ \alpha}^{E} \frac{\sum _S(E')}{(1- \alpha)E'} \phi (E') dE' +\frac{\delta(E)}{(1- \alpha)E_0}
\end{equation}
\newline
This only applicable if $\alpha E_0 < E< E_0$. A theoretical calculation is performed for an arbitrary system and Fig \ref{maxLogarithmic} is derived. For graphite the collision density is also measured for different neutron energy range as demonstrated by Fig. \ref{GraphiteUqm}.
\newline
\newline 
and for graphite:
\newline
\newline 
The oscillations are due to Plaezack Oscillations which is the fundamental phenomenon associated with the neutron slowing-down \cite{Yousry}. 
And finally for the $M \neq 1$ with capture:
\begin{dmath}
\label{eq41}
F(E)= (\sum_s (E) + \sum_a(E)) \phi(E)= \left(\int_{E/ \alpha}^{E} \frac{\sum _S(E') \phi(E')}{(1- \alpha)E'} dE' + \frac{S_0}{(1- \alpha)E_0} \right)
          = \int_{E/ \alpha}^{E} \frac{\sum _S(E')}{\sum _t(E')} \frac{F(E')}{(1- \alpha)E'} dE'+ \frac{S_0}{(1- \alpha)E_0} 
\end{dmath}

\section{Model Set-up in MCNP} 
\begin{figure}
 \begin{center}
 \begin{minipage}{3.5\textwidth}
  \ifpdf
      \includegraphics[height=4.5in]{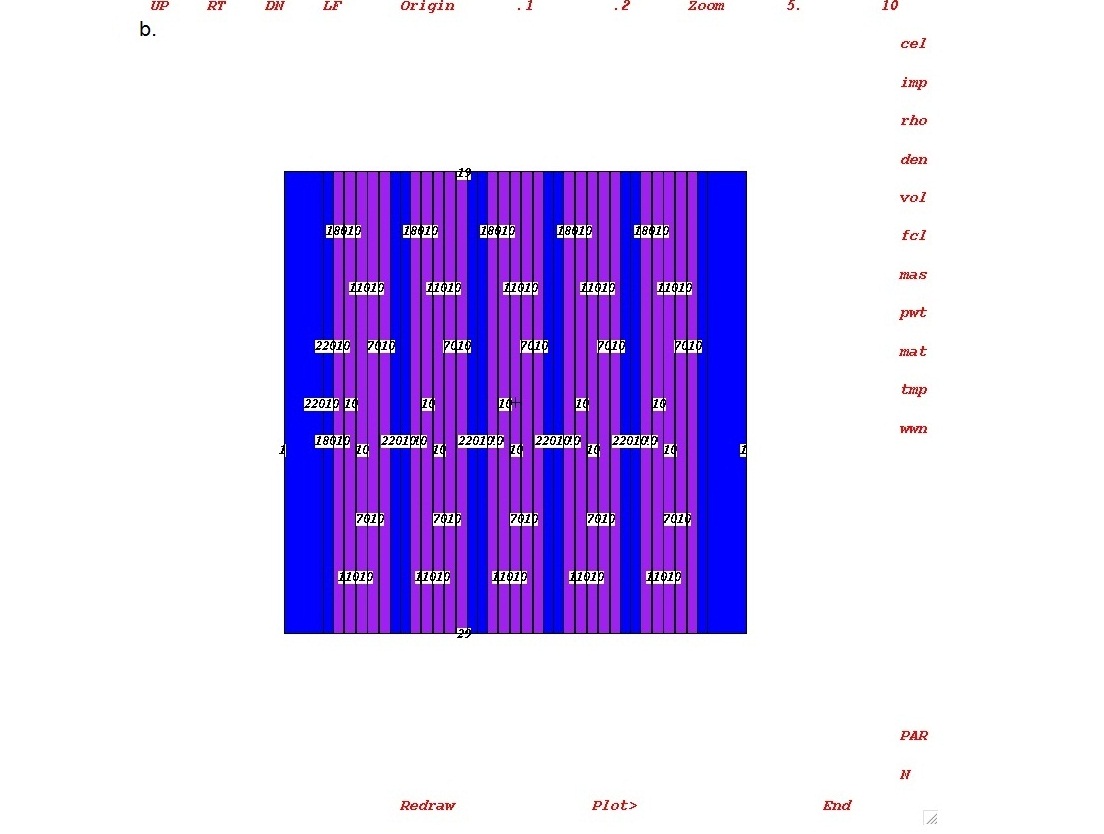}
    \else
      \includegraphics[width=0.5 \textwidth]{fuelrodins}
    \fi
 \end{minipage}
 \caption[The MCNP Geometry Setup]{The MCNP Geometry Set up: Figure demonstrates the reactor moderator module. The dimension was set 2m x 2m }
\label{reactorcore}
\end{center}
\end{figure}

The MCNP code is developed in Los Alamos National Laboratory and it is well-known for analysing the transport of neutron and $\gamma$-rays in matter. MCNP is a continuous energy modeller with generalized geometry time dependent code that implements data from nuclear libraries such as, Evaluated Nuclear Data File (ENDF), Evaluated Nuclear Data Library (ENDL), Activation Library (ACTL).
\newline
The code structure is divided into four main sections. geometry definitions, surface definitions, material cards, and tallies. Geometry of MCNP is a three-dimensional form defined using cell and surface cards. For instance Fig \ref{reactorcore} demonstrates the geometry setup in this system. 
Figure \ref{reactorcore}- illustrates the reactor moderator, where each cylinder represents the fuel rod containing slightly enriched $^{235}U$. The moderator is reactor grade graphite. 
\newline
The user can instruct the code to make various analysis with tally cards. The tallies are to measure the particle current on the surface, particle flux and energy deposition. In fact any quantity in form of Eq. \ref{ch3eq1} can be tallied \cite{Shultis}.
\newline
\begin{equation}
\label{ch3eq1}
C= \int \phi(E) f(E) dE
\end{equation}
\newline
Here $\phi(E)$ represent the particle flux and  $f(E)$ is the cross-section quantities given in the libraries. Table \ref{table:nonlin} demonstrates the six MCNP standard tallies.
\newline
\begin{table}[ht]
\caption{MCNP Neutron Tallies} 
\centering 
\begin{tabular}{c c} 
\hline\hline 
Property & Data \\ [0.5ex] 
\hline 
F1:N & Surface Current         \\
F2:N & Surface Flux         \\
F4:N & Track Length Estimate of Cell Flux         \\
F5a:N & Flux at a Point\\ 
F6:N &  Track Length Estimate of Energy Dependence        \\
F7:N & Track Length Estimate of Fission Energy Dependence  \\
\hline 
\end{tabular}
\label{table:nonlin} 
\end{table}  
In MCNP when neutron collides with a nucleus: the nuclide will be identified depends on the preferences of target, that is either the $S(\alpha, \beta)$ treatment or velocity of target; therefore the nucleus will be sampled for low energy neutrons; neutron capture or absorption will be modelled and either elastic or inelastic reaction depend on the model performance.  
\newline
However sometimes different nuclide form a material, (where the collision occurs) therefore we can have:
\newline
\begin{equation}
\label{ch3eq2}
\sum_{i=1}^{k-1} \sum_{ti} < \xi \sum_{i=1}^{n} \sum_{ti}  \leq \sum_{i=1}^{k} \sum_{ti} 
\end{equation}
\newline
Where $\sum_{ti}$ is the microscopic total cross-section of nuclide $i$. The total cross-section is sum of the capture cross-sections in the cross-section reference table. 
\newline
The collision between thermal neutrons and the target will be effected by thermal motion of the atoms, chemical binding and lattice structure of the target. This is called Free Gas Thermal Treatment. Hence the effective scattering cross-section in laboratory system is given by \cite{RICC}:  
\newline
\begin{equation}
\label{ch3eq3}
\sigma_s^{eff} (E)= \frac{1}{v_n} \int \int \sigma_s(v_{ref})v_{rel}P(v) dv \frac{d \varphi_t}{2}
\end{equation}
\newline
Here $v_n$ is particle velocity, $v_{rel}$ is the relative velocity, $P(v)$ is the probability density function and $\varphi$ as explained before is the cosine angle of velocity vector. The relative velocity can therefore is given by: 
\newline
\begin{equation}
\label{ch3eq4}
v_{rel} =(v^2_n +v^2 -2v_n v \varphi_t )^{1/2}
\end{equation}
\newline
The density function is also given by:
\newline
\begin{equation}
\label{ch3eq5}
P(v)= \frac{4}{\pi^{1/2}} \beta^3 v^2 e^{- \beta^2} v^2
\end{equation}
\newline
where $\beta=(\frac{AM_n}{2kT})^{1/2}$. However most of the time in equation \ref{ch3eq3} the $\sigma_s$ can be ignored for heavy nuclei, where $\sigma_{rel}$ can have moderating effect and is given by \cite{RICC}:
\newline
\begin{equation}
\label{ch3eq5}
P(v, \varphi) \propto \sqrt{v^2_n v^2-2v.v_n \varphi_t v^2 e^{-\beta^2 v^2}}
\end{equation}
\newline
In MCNP there are also two types of capture, analogue and implicit. Analogue occurs when the particle is killed with probability of $\frac{\sigma_a}{\sigma_t}$. Where $\sigma_a$ and $\sigma_t$ is the absorption and total cross-section respectively. Implicit capture happens when neutron weight ($W_n$) is reduced by number of collisions and is given by:
\newline
\begin{equation}
\label{ch3eq6}
W_{n+1}=(1-\frac{\sigma_a}{\sigma_t} )W_n
\end{equation}
\newline
The elastic scattering directed by two body kinematics:
\newline
\begin{equation}
\label{ch3eq7}
E_{out}=\frac{1}{2} E_in ((1-\alpha) \Theta_cm +1+ \alpha)
\end{equation}
\newline 
Where $\Theta_cm$ is the center of mass cosine of angle between incident and existing path direction. Where in inelastic an scatter the particle reaction is chosen such as $(n,n')$, $(n,2n)$, $(n,f)$, and $(n,n'\alpha)$, and is given by \cite{RICC}:
\newline
\begin{equation}
\label{ch3eq8}
E'=E'_cm+(\frac{E+2 \Theta (A+1)\sqrt{EE'_{cm}} }{(A+1)^2})
\end{equation}
\newline 
and 
\newline
\begin{equation}
\label{ch3eq9}
\varphi_{lab}=\Theta_{cm} \sqrt{\frac{E'_{cm}}{E'}} +\frac{1}{A+1}\sqrt{\frac{E}{E'}}
\end{equation}
\newline 
Here $\varphi_{lab}$ is cosine of laboratory scattering angle. However for thermal energy neutron, $S(\alpha, \beta)$ treatment is needed. For inelastic treatment the secondary particle distribution will be represented by set of discrete energies between $4eV$ to $10^{-5}eV$. 

\section{Model Set-up in COMSOL}
The Partial Differential Equation (PDE) module of COMSOL package supports three types of formation: coefficient form, general form, weak form. The coefficient form is a linear system where as the general and weak form supports non-linear, and more flexible form of definitions is supported by weak form. In this report one study is performed for thermal group transport using equation based general form of the system.
\newline
In equation based system the independent variable $u_1, u_2$ will be defined in following equation:
\newline
\begin{equation}
\label{ch3eq10}
e_a \frac{\delta^2 u}{\delta t^2}+d_a \frac{\delta u}{\delta t} - \bigtriangledown.(c \bigtriangledown u+ \alpha u - \gamma)+\beta. \bigtriangledown u+au=f
\end{equation}
\newline 
Where $e_a$ is the mass matrix, and $e_a \frac{\delta^2 u}{\delta t^2}$ is called mass term. $d_a \frac{\delta u}{\delta t}$ is called damping term, $\bigtriangledown.(c \bigtriangledown u+ \alpha u - \gamma)$ is called diffusive flux, $\beta$ is convection flux, $a$ is the absorption coefficient, and $f$ is the source term. 
\newline
Environmental factors are defined by enforcing boundary conditions using Dirichlet equation. Dirichlet imposes Laplace equation (our transport equation) to the system domain. It is therefore more convenient to have the numerical Laplace such that:
\newline
\begin{equation}
\label{fluxe}
\bigtriangledown  = \frac{\delta^2 U}{\delta x^2}+ \frac{\delta U}{\delta y^2}
\end{equation}
\newline
Now that Laplace equation is defined we need to numerically define the flux and multiply the two values, hence:
\newline
\begin{equation}
\bigtriangledown .(-c \bigtriangledown u -\alpha u+ \gamma )
\label{fluxeq3.2}
\end{equation}
\newline
Equation \ref{fluxeq3.2} is called flux vector. Here $ \alpha$ is the velocity term, $\gamma$ is the source term. $c$ can be also indirectly calculated for an anisotropic material.
\newline
Equation \ref{ch3eq10} is in computational domain ($ \omega $), thus the calculations need to satisfy all conditions in boundary domain. This is called Neumann-Dirichlet where the boundary will be transformed from $\omega$ to $d \omega$ (from computational boundary to domain boundary). This transformation is also described as domain decomposition preconditioner \cite{widlund}. Thus the partial differential equation is given by $ \bigtriangledown^{2} u+u=0 $ where $ \bigtriangledown $ donated as  Laplacian therefore we will have:
\newline
\begin{equation}
\label{emc}
\frac{du}{dn} (x)= \bigtriangledown^2 u(x). n(x)
\end{equation}
\newline
Where $n$ refers to a normal vector, thus we can rewrite the equation \ref{fluxeq3.2} as:
\newline
\begin{equation}
\label{emc}
 n .(c \bigtriangledown u + \alpha u- \gamma )= g -h^t_N
\end{equation}
\newline
where $g$  and $h^t_N$ donate the boundary source term and the Lagrange multiplier factor. $h^t_N$ is needed in a mixed field situation as it corresponds to local maxima and minima. In some respect $h^t_N$ can also refers to the velocity \cite{Babuska}.
\newline
Now by taking the energy dependent diffusion equations we have: 
\newline
\begin{dmath}
\label{dde}
\frac{1}{v} \frac{\delta \phi}{\delta t}. \bigtriangledown D \bigtriangledown \phi + \sum_t \phi= \left. \int_{0}^{\infty} \sum_s ( E \to E') \phi (E') dE' \right.
          +\chi(E) \int \bar{v}(E') \sum_f (E') \phi(E) d(E') +S(r,E,t)
\end{dmath}
Where in multi-group theory discrete energies varies with G discrete group as:
\newline
\begin{figure}[!htbp]
  \begin{center}
    \leavevmode
    \ifpdf
      \includegraphics[height=1in]{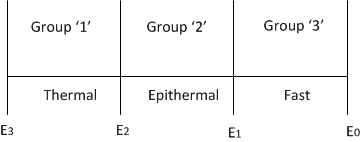}
    \else
      \includegraphics[width=0.5 \textwidth]{disgroup}
    \fi
    \caption[Discrete Group Relation, Thermal, Epithermal, Fast Region ] {Discrete Group Relation, Thermal, Epithermal, Fast Region }
    \label{heat}
  \end{center}
\end{figure}
\newline 
The group flux can be obtained by integrating total fluxes across the group energy range. Hence the parameters can be defined as below:
\newline
I. Total Cross Section: 
\newline
\begin{equation}
\label{emc}
\sum_{t}^{g} \phi_g= \int_{E_g}^{E_g-1} \sum_t (E) \phi(E) dE= \int_{E_g}^{E_g-1} \frac{\sum_t(E) \phi(E)}{\phi_g} dE
\end{equation}
\newline 
II. Diffusion Length:
\newline
\begin{equation}
\label{emc}
D_g= \frac{\int_{E_g-1}{E_g} D(E)\bigtriangledown  \phi(E) dE}{\int_{E_g-1}{E_g} \bigtriangledown  \phi(E) dE}
\end{equation}
\newline 
III. Inverse Velocity:
\newline
\begin{equation}
\label{emc}
\frac{1}{v_g}=\frac{\int_{E_g-1}{E_g} \frac{1}{v(E)} \phi(E) dE}{\phi_g}
\end{equation}
\newline 
IV. Fissile Spectrum Term
\newline
\begin{equation}
\label{emc}
\chi_g= \int_{E_g-1}{E_g} \chi(E)dE
\end{equation}
\newline 
Therefore the stationary solution for many group equation can be given by(in this work the equation is only solved for thermal group spectrum):
\newline
\begin{equation}
\label{emc}
\frac{1}{v_g} \frac{\delta \phi_g}{\delta t}. \bigtriangledown D_g \bigtriangledown \phi_g + \sum_t \phi_g=  \sum_{g'=1}^{G} \sum_s^{g' \to g} \phi_g'+\chi_g  \sum_{g'=1} (G) \bar{v}_{g'}\sum_f^{g'} \phi_g + S_g
\end{equation}
\newline
Where the right and left hand side of the equation present the loss and production term respectively, $v$ is the average neutron speed $\chi_g$ is the fraction of prompt neutrons. For the simulation the Arbitrary Lagrangian Eulerian (ALE) mapping mesh analysis is used. 

\section{Discussion and Results}
\begin{figure}
 \begin{center}
 \begin{minipage}{3.5\textwidth}
  \ifpdf
      \includegraphics[height=3.7in]{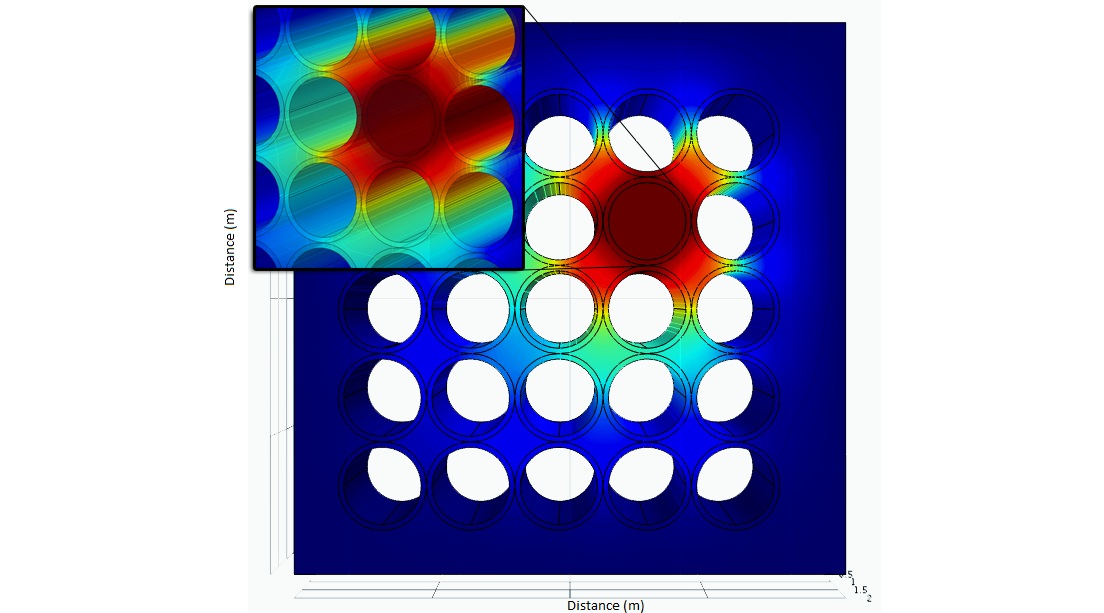}
    \else
      \includegraphics[width=0.5 \textwidth]{twoinone2}
    \fi
 \end{minipage}
 \begin{minipage}{3.5\textwidth}
  \ifpdf
      \includegraphics[height=3.5in]{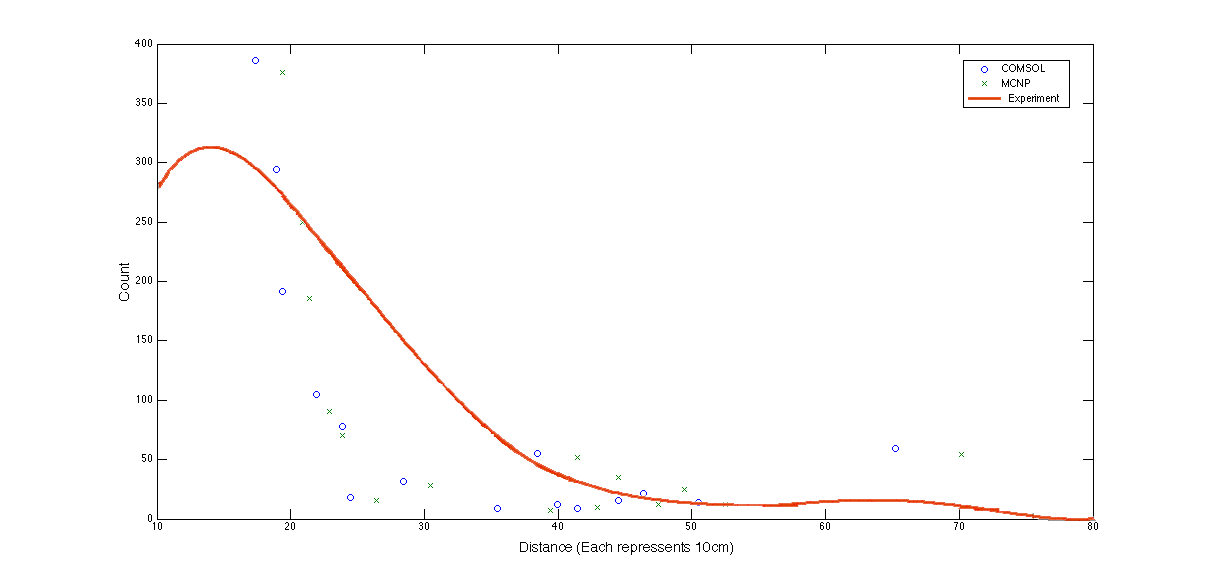}
    \else 
      \includegraphics[width=0.5 \textwidth]{mcnpvscomsol}
    \fi
 \end{minipage}
 \caption[The Neutron Diffusion Length in Graphite Assembly using FEA and COMSOL]{Figure demonstrates the the neutron diffusion length using COMSOL and MCNP, the circles illustrates the COMSOL results where as the stars demonstrate MCNP. The red line as well shows the experiment was done using BF3 tube.}
\label{commcn}
\end{center}
\end{figure}

This report has reviewed the neutron diffusion length both using COMSOL Multiphysics and MCNP. The total number of 5,000,000 meshes used for the iteration process in COMSOL. Both the thermal neutron flux and absorption property of graphite with respect to its cross-section features have been evaluated. The thermal diffusion length therefore calculated was $50.85 \pm 0.3cm$ in COMSOL and $50.95 \pm 0.5cm$ in MCNP. Figure \ref{commcn} demonstrates the distribution of thermal neutron increases as they penetrate deeper into graphite compared both in MCNP and COMSOL. The red line in the figure is also illustrates the experiment done in the lab on graphite using $Am-Be$ source. The $Am-Be$ was canned on top of an aluminium cylindrical tube. Two set-up is used in this experiment, a cadmium cover with nominal thickness of 0.1 cm (As explained previously the cadmium has cut-off of $0.55eV$) and were constructed to fully overlap the detector edges to avoid leakages. The flux distribution is measured by putting the source at a fixed location and relocating the detector at 25 cm distance intervals in horizontal and vertical directions.    

\newpage
As shown in figure \ref{commcn} the distribution calculated using COMSOL is less than half order of magnitude higher than MCNP.  For completion the absorption probability cross-section in FEA is also evaluated using inverse iteration technique as demonstrated by figure \ref{Absorption}. It is clearly shown as the neutron travels deeper into graphite they probability of absorption in graphite is also increases.
 \newline 
\begin{figure}[!htbp]
  \begin{center}
    \leavevmode
    \ifpdf
      \includegraphics[height=3.5in]{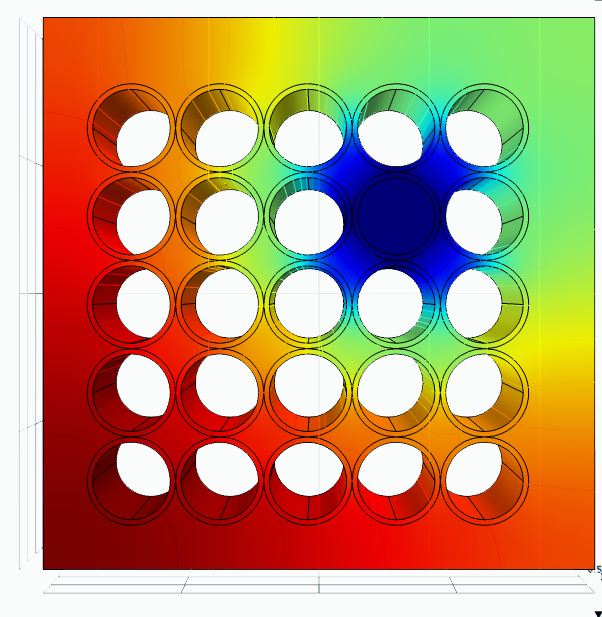}
    \else
      \includegraphics[width=0.5 \textwidth]{Absorption}
    \fi
    \caption[The Total Neutron Absorption in Graphite]{The Total Neutron Absorption in Graphite: Derived Using COMSOL. It shows the probability of absorption increases as the neutrons travel deeper in to the graphite, the areas red shows the highest and blue the lowest.}
    \label{Absorption}
  \end{center}
\end{figure}
To understand the respond of absorption cross-section to different thermal neutron energies, the evaluated values are compared with with \cite{ike}, \cite{ikeI}, \cite{ornl}, \cite{ncsu}, \cite{Neill}, \cite{Steyerl} and \cite{Palevsky} as demonstrated in figure \ref{Absorption}. 
 \newline
\begin{figure}[!htbp]
  \begin{center}
    \leavevmode
    \ifpdf
      \includegraphics[height=3.5in]{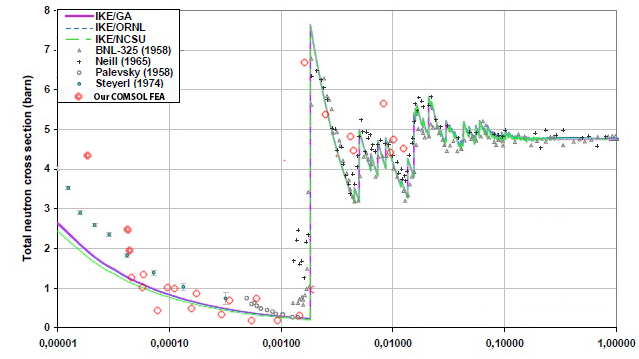}
    \else
      \includegraphics[width=0.5 \textwidth]{totalcross}
    \fi
    \caption[The Total Neutron Absorption Cross-section in Graphite Compared]{The Total Neutron Absorption Cross-section in Graphite Compared with \cite{ike}, \cite{ikeI}, \cite{ornl}, \cite{ncsu}, \cite{Neill}, \cite{Steyerl} and \cite{Palevsky}. Fittings are extracted by permission from \cite{Wolfgang}}
    \label{Absorption}
  \end{center}
\end{figure}
\newline

\section{Acknowledgement}
The author is grateful to Birmingham University colleague and professors
for stimulating discussions. Computations were performed
in the Nano Laboratory of the Department of Mechanical engineering and Department of Physics and Astronomy at the University of Birmingham. 

\newpage
\section*{References}
\bibliography{iopart-num}


\end{document}